\documentclass[iop,onecolumn,apj]{emulateapj}
\usepackage{epsfig}

\newcommand{\be}{\begin{equation}}
\newcommand{\ee}{\end{equation}}
\newcommand{\ba}{\begin{eqnarray}}
\newcommand{\ea}{\end{eqnarray}}

\newcommand{\rb}{\bar{r}}

\newcommand{\bigo}{\mathcal{O}}

\newcommand{\bOmega}{\bar{\Omega}}

\newcommand{\bnabla}{{\bf \nabla}}
 
\slugcomment{Submitted to ApJ}
\shorttitle{Acceleration Due to Gravity}

\begin{document}

\title{Universality of the Acceleration Due to Gravity on the Surface of a
Rapidly Rotating Neutron Star}
\author {Mohammad AlGendy\altaffilmark{1} , \& Sharon M. Morsink\altaffilmark{1,2}}

\altaffiltext{1}{Department of Physics, University of Alberta, Edmonton AB,  T6G~2E1, Canada}
\altaffiltext{2}{Steward Observatory, Department of Astronomy, University of Arizona, 933 N. Cherry Ave., Tucson AZ 85721, USA}

\begin{abstract}
On the surface of a rapidly rotating neutron star, the effective centrifugal force decreases the
effective acceleration due to gravity (as measured in the rotating frame) at the equator
while increasing the acceleration at the poles due to the centrifugal flattening of the star
into an oblate spheroid. We compute the effective gravitational acceleration for relativistic
rapidly rotating neutron stars and show that for a star with mass $M$, equatorial radius $R_e$,
and angular velocity $\Omega$,
the deviations of the
effective acceleration due to gravity from the nonrotating case take on a universal form
that depends only on the compactness ratio $M/R_e$, the dimensionless square of the 
angular velocity $\Omega^2R_e^3/GM$,
and the latitude on the star's surface.
This dependence is universal, 
in that it has very little dependence on the neutron star's equation of state. 
The effective gravity is expanded in the slow rotation limit to show the 
dependence on the effective centrifugal force, oblate shape of the star and the quadrupole 
moment of the gravitational field. In addition, an empirical fit and simple formula for the 
effective gravity is found. We find that the increase in the acceleration due to gravity at the poles is of the same 
order of magnitude as the decrease in the effective acceleration due to gravity at the equator 
for all realistic value of mass, radius and spin.  For neutron stars that spin with frequencies 
near 600 Hz the difference between the effective gravity at the poles and the equator 
is about 20\%. 
\end{abstract}

\keywords{stars: neutron  --- stars: rotation  --- relativity
--- pulsars: general }

\section{Introduction}
\label{s:intro}

Slowly rotating neutron stars have properties that show a surprising universality that appear to be independent of the equation of state (EOS): given a neutron star's moment of inertia, a simple formula then determines the star's quadrupole moment and tidal Love number \citep{Yagia,Yagib}. This type of universal relation has been extended to the star's ellipticity \citep{Baubock2013} and to rapidly rotating neutron stars \citep{Doneva2014,Yagi2014}. These ``I-Love-Q" relations have only an implicit dependence on the EOS: the EOS limits the possible values of mass and radius that a star with given spin may have. Once the mass, radius and spin are determined, the other properties are determined through the universal relations. In other words, the EOS limits the values the star's moment of inertia may 
take, but once this is known, the I-Love-Q relations determine the Love number and quadrupole moment. 

In this paper, we derive equations for the effective acceleration due to gravity on the surface of a rapidly rotating neutron star that
 only have an implicit dependence on EOS. The effective acceleration due to the gravity, also known as the effective gravity, 
is an important property in astrophysics, and is required in 
many applications, such as atmosphere modeling \citep{Heinke06,Sul},  X-ray bursts \citep{Spitk,Cooper}, 
and Eddington limited X-ray bursts \citep{Ozel2013}. The results in this paper can be applied to any neutron star spinning with a
frequency less than the break-up limit, however the results are most relevant for the stars that spin with frequencies above approximately
500 Hz. There are at least 15 accretion and rotation powered neutron stars with spins in this range \citep{Watts,Papitto2014}.

The I-Love-Q relations could be derived by first noting that the expressions for moment of inertia $I$, Love number and quadrupole moment $q$ depend only (after spin is factored out) on the dimensionless compactness ratio, defined by
\be
x = \frac{M}{R_e}
\ee
where $R_e$ is the radius of the star's equator as measured using the Schwarzschild radial coordinate. We use geometric units where $G=c=1$. If we have a function for the moment of inertia of the form $I(x)$, this can (assuming good mathematical behaviour) be inverted to form an equation for the compactness of the form $x(I)$. If we have an equation for $q$ that is also of form $q(x)$,
then this is equivalent to the universal form $q(I)$. In this way, a formula for any quantity that depends only on the compactness can be thought of as universal.  The introduction of rapid rotation requires the introduction of another dimensionless ratio, the dimensionless angular velocity $\bOmega$,
\be
\bOmega = \Omega \left(\frac{R_e^3}{M}\right)^{1/2}.
\ee
Properties of a rotating star that depend only the two dimensionless ratios are universal in the following sense: given an EOS, the possible values of the two ratios are determined. Once values for $x$ and $\bOmega$ are known, then the property of the star is known with only an implicit dependence on the EOS. This type of universality has previously been noted for the star's oblate shape \citep{Morsink2007}, and the universality has been an important feature allowing the extraction of
mass-radius constraints for accretion-powered pulsars \citep{Morsink2011,Leahy2011}.

 In this paper, we derive the dependence of the effective gravity on the two dimensionless parameters and show that it too has a universal nature that is almost
independent of the equation of state. This universality allows us to find a simple empirical formula for the effective gravity.

In Section \ref{s:rotate} we introduce the metric for rotating stars, as well as the slow rotation expansion of the metric. The parametrized expansion that we use to construct universal relations is introduced in Section \ref{s:para} and applied to the moment of inertia, quadrupole moment and stellar oblateness, since these quantities will be required in the computation of the effective gravity. The derivation of the formula for the effective gravity is made in Section \ref{s:accel} 
and universal nature of its dependence on the dimensionless parameters demonstrated. In Section \ref{s:slow}, the slow rotation expansion is applied to the 
formula for the effective gravity in order to aid the intuition in how it depends on the moment of inertia, quadrupole moment and oblate shape. In Section
\ref{s:rapid} we derive a simple empirical formula for the effective gravity on the surfaces of rapidly rotating stars. Finally, we conclude in Section \ref{s:final}
with a discussion of astrophysical applications.

\section{Rotating Relativistic Stars}
\label{s:rotate}

The metric for a stationary, axisymmetric rotating star can be written using the metric \citep{Butterworth1976}
\be
ds^2 = -e^{2\nu} dt^2 + \rb^2 \sin^2 \theta B^2 e^{-2\nu} (d\phi - \omega dt)^2
+ e^{2\zeta-2\nu} (d\rb^2 + \rb^2 d\theta^2),
\label{eq:BImetric}
\ee
where metric potentials $\nu, B, \zeta,$ and $\omega$ are functions of the coordinates $\rb$ and co-latitude $\theta$.  A circle around the star in the equatorial plane has circumference $2 \pi r$, where the circumferential radius $r$ is related to the coordinate $\rb$ by \citep{FIP}
\be
r = B e^{-\nu} \rb.
\ee
In the limit of zero rotation, the coordinate $r$ reduces to the Schwarzschild radial
coordinate, while $\rb$ reduces to the isotropic Schwarzschild radial coordinate,
\be
\lim_{\Omega \rightarrow 0} r = \rb \left( 1 + \frac{M}{2\rb}\right)^2.
\ee
For more details about the coordinates and metric for rapidly rotating neutron stars see \citet{FS2013}.

Given a perfect fluid equation of state, the Einstein field equations
can be solved numerically \citep{CST94} for the metric potentials. All computations in this paper are done using the Rapidly Rotating Neutron Star (RNS) 
code\footnote{ Code available at \url{http://www.gravity.phys.uwm.edu/rns/}}
 \citep{SF}. We made use of a very fine computational grid with 301 angular divisions and 801 radial divisions in order to obtain good 
precision in the values for the stars' parameters (such as the quadrupole moment).

\subsection{Slow Rotation Approximation}

The \citet{Hartle1967} slow rotation approximation is an expansion of the metric in powers of $\bOmega$, keeping terms up to order $\bOmega^2$. For the purpose of this paper, we consider values of $\bOmega^2 < 0.1$ to represent slow rotation, although it is not incorrect to make use of the slow rotation approximation for higher values of $\bOmega^2$. For realistic equations of state, most neutron stars with mass ($M>1.0 M_\odot$) and spin ($\nu_\star < 700$ Hz) have $\bOmega^2 \le 0.1$. 

In the limit of zero spin, the \citet{Butterworth1976} metric reduces to the isotropic 
Schwarzschild metric. This limit will be denoted with the subscript ``$0$" so that the
metric potentials are
\begin{eqnarray}
e^\nu_0 &= & \left( 1-\frac{M}{2\rb}\right) \left( 1+\frac{M}{2\rb}\right)^{-1} 
	= \left(1 - \frac{2M}{r}\right)^{1/2}  \label{eq:nu_0}\\
B_0 = e^{\zeta_0} &= & \left( 1-\frac{M}{2\rb}\right) \left( 1+\frac{M}{2\rb}\right)\\
\omega_0 &= & 0. \label{eq:omega_0}
\end{eqnarray}

The lowest order terms in the series expansions for the metric potentials are
\citep{Butterworth1976,FS2013}
\begin{eqnarray}
\nu &=&  \nu_0 + \bOmega^2 \nu_2 =
\nu_0  + (\frac{\beta}{3} - q P_2(\cos\theta)) \left( \frac{M}{\rb} \right)^3 
+ \bigo\left( \bOmega^2 \times \left(\frac{M}{\rb} \right)^4 \right)\\
B &=& B_0  +  \beta \left( \frac{M}{\rb} \right)^2 + \bigo(\bOmega^4) \times \bigo\left( \frac{M}{\rb} \right)^4\\
\omega &=& \frac{2j}{\rb} \left( \frac{M}{\rb} \right)^2 \left( 1 - 3 \left( \frac{M}{\rb} \right)  +  \bigo\left( \frac{M}{\rb} \right)^2 \right) + \bigo(\bOmega^3) \\
\zeta &=& \zeta_0 + \bar{\Omega}^2 \zeta_2 = \ln(B_0) +  \beta \left( \frac43 P_2(\cos\theta) - \frac13 \right) 
\left( \frac{M}{\rb} \right)^2 +  \bigo( \bOmega^2 ) \times \bigo\left( \frac{M}{\rb} \right)^4, \label{eq:zeta}
\end{eqnarray}
where the leading order terms in the post-Newtonian expansions for $\nu$, $\omega$, and $\zeta$ are displayed. 
The dimensionless constants $j, q$, and $\beta$ characterize the solution. The dimensionless quantity $j$ is the ratio $J/M^2$, where $J$ is the star's angular momentum. The dimensionless quantities $q$ and $\beta$ are the dimensionless moments of the energy density and pressure and are related to the Butterworth-Ipser quantities $\tilde{\nu}_2$ and $\tilde{B}_0$  by $q = -\tilde{\nu}_2/M^3$ and $\beta = \frac14 + \tilde{B}_0/M^2$. Both $q$ 
and $\beta$ are $\bigo(\bOmega^2)$, while $j$ is $\bigo(\bOmega)$.

\section{ Parametrized Expansions for Neutron Star Properties}
\label{s:para}

Many properties of a relativistic star follow universal formulae \citep{Yagib, Baubock2013, Doneva2014,Pappas2014} that do not 
explicitly depend on the equation of state. Instead, given a value for the star's moment of inertia, the quadrupole moment is given by a simple formula that is (within small error limits) independent of the equation of state. This is equivalent to stating that the star's properties 
(such as moment of inertia or quadrupole moment) only depend explicitly 
on the
star's compactness $x = \frac{M}{R}$ and the dimensionless angular velocity $\bOmega$, 
as was found earlier for the star's oblate shape \citep{Morsink2007,Baubock2013}. There 
is still a dependence on the EOS, but this dependence is implicitly defined: the EOS 
determines what values of $M$ and $R$ are allowed given an angular velocity $\Omega$. 
Once $M$, $R$, and $\Omega$ are known, the dimensionless parameters $x$ and $\bOmega$
are determined.  

The neutron star's properties can be split into primary and secondary properties. The primary
properties are the mass, equatorial radius, and the angular velocity. Any equation of state provides a unique relation between the three primary properties. From the 3 primary properties, the two dimensionless ratios $x$ and $\bOmega$ can be constructed. Given an EOS, there is then a range of possible values for the two dimensionless ratios. The secondary properties, such as moment of inertia, Love number, quadrupole (and other) moments, oblateness, acceleration due to gravity, etc, have a dependence only on $x$ and $\bOmega$.  A similar universal relation has been found for the orbital frequency of a particle at the innermost stable
circular orbit \citep{Kluz1990}, although they express the frequency in terms of the angular momentum (or equivalently moment of inertia) instead of $\bOmega$.
Calculations of the mass-shed limit have shown that a universal formula can predict the fastest allowed spin frequency as well \citep{Haensel,FIP89,Lasota}. However, the dependence of 
the maximum spin frequency on $x$ and $\bOmega$ is not directly shown in these formulae.

The dependencies of these secondary properties on the two dimensionless ratios do depend on the EOS, but only very weakly, so that it is possible to create very good approximate formulae for the secondary properties. The approximate formulae for the secondary properties are found by (1) choosing a library of representative EOS; (2) computing neutron star models numerically for a reasonable range of mass, radius and spin values for the EOS; (3) computing the dimensionless ratios and the desired secondary properties; and (4) finding the best fit approximations for each property. 

For our approximate formulae, we choose a library of representative EOS that span the range of stiffness allowed by observations. The approximate formulae do depend on the choice of EOS, but the errors introduced are small. We explore this dependence in our computations for the moment of inertia. Our library of EOS includes the following: a hadronic EOS (BBB2) \citep{BBB};
an EOS (APR) making use of an effective 3-particle scattering potential and first order special relativistic corrections \citep{APR98}; a hybrid quark-hadronic EOS (ABPR1) \citep{ABPR};
 a hyperon (H4) EOS \citep{Lackey}; and
the set of softest (HLPS1), 
mid-range (HLPS2), and stiffest (HLPS3) EOS consistent with  modern nuclear interaction theory, computed by \citet{HLPS}. All of these EOS meet the requirement that they allow
a 1.93 $M_\odot$ neutron star, as observed by \citet{Demorest}. 
For each EOS, we compute models with masses between $1.0 M_\odot$ and the maximum mass star allowed by the EOS. Rotation rates up to the mass-shed limit are computed. However, for the parametrized fits given in this section, only models with $\bOmega^2\le 0.1$
are used. The total number of stellar models computed is on the order of 39,000 with approximately 5,000 models having parameters in the slow rotation 
range. We restrict the fits (in this section) to the slow rotation limit, since many of the parameters, 
such as $q$ and $\beta$ are defined only through an expansion in $\Omega$. In addition, we require a consistent set of fits for these parameters in the slow rotation limit in order
to compare with the acceleration in the same limit.

The star's moment of inertia $I$ is parametrized by 
\be
I = i(x,\bOmega) MR_e^2, 
\label{eq:idefn}
\ee
where $i(x,\bOmega)$ is a dimensionless function to be determined by a least-squares fitting procedure. Making
use of the definition of angular momentum,
the relation between $j$ and $\Omega$ is 
\be
j =  \frac{J}{M^2} = i(x,\bOmega)  \Omega M \left( \frac{R_e}{M}\right)^2 = i(x,\bOmega)  \bar{\Omega} \sqrt{ \frac{R_e}{M}}.
\label{eq:j}
\ee

We find that the simplest consistent fit for the moment of inertia is the function
\be
i(x,\bOmega) = x^{1/2}( i_0 + i_1 x + i_2 x^2  ),
\label{eq:inertia}
\ee
consistent with the results of \citet{Baubock2013} who fit the moment of inertia using the slow rotation approximation. The $i_n$ coefficients are shown in Table \ref{tab:coeff}. An error estimate, the RMS summed square of the residuals, labelled as $\sigma$, is given in Table \ref{tab:coeff}, defined as
\be
\sigma = \left( \frac{1}{N-n} \Sigma_{i=1}^{N} (i(x,\bOmega) - i_i)^2   
\right)^{1/2},
\ee
where $i(x,\bOmega)$ is the model given in Equation (\ref{eq:inertia}), $i_i$ is the moment of inertia of the $i$th stellar model,  $N$ is the number of stellar models and $n$ is the number of free parameters in the fit. The error in computing the moment of inertia for a particular model is very small: when we double the computation grid used by RNS, the change in $i_i$ is less than 0.01\%. However, when we make use of the model in Equation (\ref{eq:inertia}) we 
are assuming that all EOS can be described by exactly the same set of parameters. The standard deviation $\sigma$ is a measure of how much the 
different EOS deviate from the universal form, and we expect that $\sigma$ will be larger than the computational error in calculating one star's parameters.
For the moment of inertia fit, $\sigma = 0.01$, while typical values of the dimensionless moment of inertia are approximately $0.4$, leading to about 3\% 
error in computing the moment of inertia using the empirical formula. 

\begin{deluxetable}{llrrrrl}
\tabletypesize{\small}
\tablecaption{Expansion Coefficients for Slow Rotation ($\bOmega^2 \le 0.1$)}
\tablehead{ \colhead{} & \colhead{Equation}& 
 \colhead{$a_o$} & \colhead{$a_1$} & \colhead{$a_2$}   & \colhead{$\sigma$} & \colhead{Comments}}
\startdata
 $i$ & $x^{1/2}( a_0 + a_1 x + a_2 x^2)$ & $1.1035 $ & -2.146 & 4.5756 & &\citet{Baubock2013}\\
 $i$ & $x^{1/2}( a_0 + a_1 x + a_2 x^2)$ & $1.136 \pm 0.6\%$ & $-2.53 \pm  2.3\%$  & $5.6 \pm  2.1\%$ & 0.01 & Equation (\ref{eq:inertia}) \\
$q$ & $\bOmega^2 (  {a_2}/{x^2} )$     & & & $-0.11 \pm 0.1\%$                           & 0.01& Equation (\ref{eq:quad}) \\
$\beta$ & $a_1 \bOmega^2 x $               &             &   $0.4454 \pm 0.1\%$  &&     0.0005          & Equation (\ref{eq:beta})\\
$o_{2}$& $ \bOmega^2 ( a_{0} + a_{1} x)$ & $-0.788 \pm 0.2\%$  &  $1.030 \pm 0.5\%$ &        & 0.001  &Equation (\ref{eq:oblate}) \\
\enddata
\label{tab:coeff}
\end{deluxetable}

Each coefficient in Table \ref{tab:coeff} has an error estimate, expressed as a percent of the coefficient value. This fit-coefficient percent error is derived from the
value of $\sigma$, so it is not an independent error estimate. 
The values of the coefficients computed by \citet{Baubock2013} and the present paper
agree within 5\% of each other. The 5\% difference is mainly due to the different choices of EOS used in the approximation. However, since the
coefficients are multiplied by small numbers, the actual difference in the inertia values are much smaller. For instance consider a model with $x=M/R_e=0.1$. 
If the dimensionless moment of inertia is computed with both sets of empirical fitting coefficients, the difference between the computed values is 
only 0.5\%, which is surprising good considering that 
 a different set of EOS were used in the two computations.  

One expects there to be a dependence of the moment of inertia on the dimensionless angular velocity, since the star's equatorial radius increases as the spin increases. However, since the radius $R$ used in our approximation is the equatorial radius of the spinning star, most of this dependence is already included in the $R^2$ dependence of equation (\ref{eq:idefn}). 

\cite{Laarakkers} computed the values  for $q$, the coordinate non-invariant 
dimensionless quadrupole moment. They found the scaling $ q = a(M,eos) j^2$,
where the parameter $a(M,eos)$ depends on mass and the equation of state.
A simple approximation for $q$ that depends only on 
$x$ and $\bOmega$ is 
\be
q(x,\bOmega) = \bOmega^2 ( \frac{q_2}{x^2} )
\label{eq:quad}
\ee
where the coefficient $q_2$ is shown in Table \ref{tab:coeff}. The meaning of the
error estimates are the same as for the moment of inertia errors.
This expression has the same leading order behaviour as suggested by the 
simple scaling found by \citet{Laarakkers}.

The coordinate invariant quadrupole moment is the combination of terms
\citep{Pappas}
\be
q_{inv}  = \left( q  + \frac43 \beta\right).
\ee
The parameter $\beta$ is generally small compared to $q$, and is related to the 
difference in circumference between circles of constant $\rb$ circling the equator
or the poles. The ratio $\beta/q$ was computed in \cite{Morsink1999} where it
was tabulated as the parameter $\zeta = 100/3 \times \beta/q$. This ratio 
is fairly small and only reaches the 10\% level for stars near the maximum allowed mass.
The simplest parametrization for the coefficient $\beta$ that we find is
\be
\beta = \beta_1 \bOmega^2 x,
\label{eq:beta}
\ee
where the coefficient $\beta_1$ is given in Table \ref{tab:coeff}.

The oblate shape of the star was calculated using a Legendre polynomial expansion by \citet{Morsink2007}. 
Here we recompute the oblate shape using a slightly different EOS library, and using an equivalent expansion 
in powers of $\cos^2\theta$. This expansion for the star's oblate shape is 
\be
R(\theta) = R_e \left( 1 - \frac{(R_e - R_p)}{R_e} \cos^2\theta \right) = R_e \left( 1 + o_2(x,\bOmega) \cos^2(\theta)  \right),
\ee
where $R_p$ is the radius of the star along the spin axis, and the expansion coefficient have the form
\be
o_{2} = \bOmega^2 ( o_{20} + o_{21} x).
\label{eq:oblate}
\ee
The $o_n$ coefficients are within 3\% of the values computed by the older \citet{Morsink2007} paper, once the transformation between Legendre polynomials and powers of $\cos^2\theta$ have been made. There are two reasons for the small differences: the older paper made use of many softer EOS that have since been ruled out; the approximation in this paper only makes use of data for slow rotation, and in the older paper we did not enforce the identity $R(\pi/2) = R_e$.

\section{Acceleration Due to Gravity on a Rotating Star}
\label{s:accel}

\subsection{Newtonian Physics}

The equation of hydrostatic equilibrium in Newtonian physics is 
${dP}/{dr} = -g \rho $
where $g$ is the acceleration due to gravity. On the surface of a non-rotating star,
$g = GM/R^2$. The generalization of the equation of hydrostatic equilibrium in a rotating frame is
$
 \bnabla P = - \rho{\bf g}
$
where the effective ${\bf g}$ is defined by \citep{Tassoul}
\be
{\bf g} = - {\bnabla} \Phi_{e} = - {\bnabla} \left( \Phi - \frac12 |{\bf \Omega} \times {\bf r} |^2 \right)
\ee
where $\Phi$ is the gravitational potential and $\Phi_e$ is the effective potential in a rotating frame. 
Since the effective acceleration due to gravity 
can be derived from a potential, the vector ${\bf g}$ is normal to surfaces of constant pressure and density (which coincide). The
magnitude of the effective acceleration on the surface of the star is then $g = {\bf n} \cdot {\bf g}$,
where ${\bf n}$ is the unit normal to the surface.

The surface of a rotating star is a function of co-latitude, $R(\theta)$, found by solving the stellar structure equations. 
For slow rotation,
the surface function has the form given by Equation (\ref{eq:oblate}).
 The angle
$\gamma$, defined to be the angle between the normal to the surface, $\bf{n}$ and the radial direction $\bf r$ is given by 
\be
\cos \gamma = \left( 1 + \left( \frac{1}{R} \frac{dR}{d\theta} \right)^2 \right)^{-1/2}.
\ee
Since departures from sphericity are of order $\bar{\Omega}^2$, it follows that $\gamma \sim \bigo(\bar{\Omega}^2)$. The normal
to the surface can be written in terms of the radial and tangential unit vectors for spherical coordinates as
$
{\bf n} = \cos\gamma {\bf r} + \sin \gamma {\bf \theta}.
$
With this definition for the normal to the surface, the effective gravity is
\be
g = - \cos\gamma \frac{\partial \Phi_e}{\partial r} - \sin\gamma \frac{1}{r} \frac{\partial \Phi_e}{\partial \theta}.
\ee
For slow rotation in Newtonian physics, the gravitational potential is
\be
\Phi(r,\theta) = - \frac{GM}{r} - \frac{G \Phi_2}{r^3} P_2(\cos\theta) + \bigo(\bar{\Omega}^4)
\ee
where $\Phi_2$ is the quadrupole moment which is $\Phi_2 \sim \bigo(\bar{\Omega}^2)$. We are using a sign 
convention where $\Phi_2$ is negative for oblate stars.

Given these definitions, for a slowly rotating Newtonian star, the effective gravity is
\be
g = \frac{GM}{R_e^2} \left( 1 - \bar{\Omega}^2 \sin^2\theta + \frac{3 \Phi_2}{M R_e^2} P_2(\cos\theta) + 2 \frac{(R_e-R_p)}{R_e} \cos^2\theta \right).
\ee
This equation for the effective gravity includes the reduction in
gravity at the equator due to the effective centrifugal force, the
gravitational quadrupole moment that increases the gravity at the
equator and reduces it at the poles, and the increase in the gravity
at the poles due to the oblate shape of the star.

\subsection{Acceleration in General Relativity}

The equation of hydrostatic equilibrium in general relativity is derived by taking the equation for the conservation of 
the energy-momentum tensor $\nabla_\beta  T^{\alpha \beta} = 0$ and projecting the equation into the subspace perpendicular to the fluid world lines \citep{FS2013}. The fluid world lines for a rotating star are described by the four-velocity $u^\alpha$
\be
u^\alpha = N (t^\alpha + \Omega \phi^\alpha) 
\label{eq:four-vel}
\ee
where $t^\alpha$ and $\phi^\alpha$ are the timelike and rotational killing vectors and the normalization function
$N$ is defined through the normalization relation $u^\alpha u_\alpha = -1$. 
Given a spacetime metric $g_{\alpha\beta}$, the tensor $q_{\alpha\beta}$, defined by
$
q^{\alpha\beta} = g^{\alpha\beta} + u^\alpha u^\beta 
$
projects tensors into the subspace perpendicular to the fluid four-velocity. The projection of the 
conservation law, $ q_{\alpha\gamma} \nabla_\beta T^{\beta \gamma} = 0$ leads to the equation of
hydrostatic equilibrium 
$ q_\alpha^\beta \nabla_\beta p = - (\rho + p) u^\beta \nabla_\beta u_\alpha = - (\rho + p) a_\alpha$
where  $a^\alpha$ is the
fluid's acceleration vector. In the case of the four-velocity (\ref{eq:four-vel}), the 
acceleration is simply 
\be
a_\alpha = - \nabla_\alpha \ln(N), \label{eq:a_alpha}
\ee
which is again, a vector normal to the star's surface \citep{FS2013}. 
The magnitude of the acceleration vector is the quantity that is identified with the 
effective acceleration due to gravity
\be
g = a = ( g^{\alpha\beta} a_\alpha a_\beta )^{1/2}. \label{eq:geff}
\ee

\subsubsection{Non-rotating Star}

In the case of a non-rotating star ($\Omega = 0$), the metric exterior to the star's surface (at radius $R$) is 
given by the Schwarzschild metric
\be
ds^2 = -\left( 1 - \frac{2M}{r} \right) dt^2 + \left( 1 - \frac{2M}{r} \right)^{-1} dr^2 + r^2 d\Omega^2.
\ee
Plugging the Schwarzschild metric into equations (\ref{eq:four-vel} - \ref{eq:geff}),
the acceleration due to gravity is 
\begin{equation}
 g_0 =\frac{M}{R^{2}}\frac{1}{(1-\frac{2M}{R})^{1/2}}.
\label{eq:g_0}
\end{equation} 
\citet{Thorne1967} derives this formula using a more intuitive approach involving the definition of buoyancy.

\subsubsection{Rotating Star}

It is convenient to define \citep{Butterworth1976} a velocity $V$ by
\be
V = \rb B e^{-2\nu} (\Omega - \omega) \sin\theta
\label{eq:vel}
\ee
where $V$ is interpreted as the 3-velocity of a fluid element in the star as measured by a zero angular
momentum observer at infinity. Given the definitions (\ref{eq:BImetric}) and (\ref{eq:vel}), the acceleration vector (\ref{eq:a_alpha}) in the Butterworth-Ipser coordinate system is
\be
a_{\alpha} 
= \frac{ \partial \nu}{\partial x^\alpha} - \frac{V}{1-V^2} \frac{ \partial V}{\partial x^\alpha}.
\label{eq:a_butter}
\ee

The coordinate-independent effective surface gravity is then
\be
g = e^{\nu - \zeta} \left( a_{\rb}^2 + \left(\frac{a_\theta}{\rb}\right)^2 \right)^{1/2}.
\label{eq:g_butter}
\ee
where all quantities are evaluated on the star's surface. This is equivalent to the 
formula used by \citet{Cumming} to compute the effective gravity at the
equator of a rotating neutron star.

A dimensionless effective surface gravity $g/g_0$ can be computed for a rapidly rotating star, where $g$ is given by 
Equation (\ref{eq:g_butter}) and $g_0$ is defined by Equation (\ref{eq:g_0}), where $R$ is the radius at the equator of the
rotating star.   A few representative models have been computed using two EOS: APR and HLPS2. 
The basic properties of the models are shown in Tables \ref{tab:HLPS2} and \ref{tab:omega}, including the value of $g/g_0$
at the equator and the poles.
Parameters were chosen so that all models have one of 3 values of the dimensionless equatorial compactness ratio $M/R_e$
and one of 4 values of the dimensionless angular velocity $\bOmega$. 
 For both EOS, the 
models with $\bOmega^2$ close to 1.0 are spinning with just a few Hz below the mass-shed limit. 
The models with $M/R_e=0.320$ are close to the maximum mass limits for both EOS. 
The models in these two tables have the property that two stars with different EOS 
but the same values of $M/R_e$ and $\bOmega^2$ 
have very similar values for the dimensionless effective surface gravity at the poles and
at the equator. Absolute differences between values for the effective gravity for two models 
are typically less than 0.03 (compared to 1.0 for a non-rotating star).

\begin{deluxetable}{lllrrrrr}
\tablecaption{Representative Neutron Star Models with $M/R_e = 0.195$}
\tablehead{ \colhead{EOS} & \colhead{$M$} & \colhead{$R_e$}& 
 \colhead{$M/R_e$} & \colhead{$\nu_\star$} & \colhead{$\bOmega^2$}  
& \colhead{$g_e/g_0$} & \colhead{$g_p/g_0$}\\
\colhead{} & \colhead{$M_\odot$} & \colhead{$km$}& 
 \colhead{} & \colhead{$Hz$} & \colhead{} & \colhead{} & \colhead{} 
 }
\startdata
HLPS2 & 1.69 & 12.8 & 0.195 & 529 & 0.10 &0.936 &1.088 \\
HLPS2 & 1.80 & 13.6 & 0.195 & 842 & 0.30 & 0.796  & 1.257\\
HLPS2 & 1.99 & 15.1 & 0.195 & 1084 & 0.60  & 0.516& 1.544\\
HLPS2 & 2.27 & 17.1 & 0.195 & 1223 & 0.99 & 0.011 & 2.019 \\
APR    &  1.55 & 11.7 & 0.195 & 568 & 0.10 & 0.935  & 1.087\\
APR    &  1.65 & 12.5 & 0.195 & 923 & 0.30 & 0.787 & 1.264 \\
APR    &  1.82 & 13.7 & 0.195 & 1190 & 0.60 & 0.502 & 1.560 \\
APR    &  2.05 & 15.5 & 0.195 & 1350 & 0.98 & 0.012 & 2.041 
\enddata
\label{tab:HLPS2}
\end{deluxetable}

\begin{deluxetable}{lllrrrrr}
\tablecaption{Representative Neutron Star Models with $\bOmega^2=0.3$}
\tablehead{ \colhead{EOS} & \colhead{$M$} & \colhead{$R_e$}& 
 \colhead{$M/R_e$} & \colhead{$\nu_\star$} & \colhead{$\bOmega^2$}  
& \colhead{$g_e/g_0$} & \colhead{$g_p/g_0$}\\
\colhead{} & \colhead{$M_\odot$} & \colhead{$km$}& 
 \colhead{} & \colhead{$Hz$} & \colhead{} & \colhead{} & \colhead{} 
 }
\startdata
HLPS2 & 1.23 & 13.4 & 0.135   & 718 & 0.30 & 0.775  & 1.283 \\
HLPS2 & 1.80 & 13.6 & 0.195 & 842 & 0.30 & 0.796  & 1.257\\
HLPS2 & 2.68 & 12.3 & 0.320  & 1199 & 0.30 & 0.812  & 1.205 \\
APR    &  1.16 &12.6  & 0.135 & 763 & 0.30 & 0.769 & 1.287 \\
APR    &  1.65 & 12.5 & 0.195 & 923 & 0.30 & 0.787 & 1.264\\
APR    &  2.37 & 10.9 & 0.320 & 1351 & 0.30 & 0.809 & 1.207
\enddata
\label{tab:omega}
\end{deluxetable}

\begin{figure}
\begin{center}
\plotone{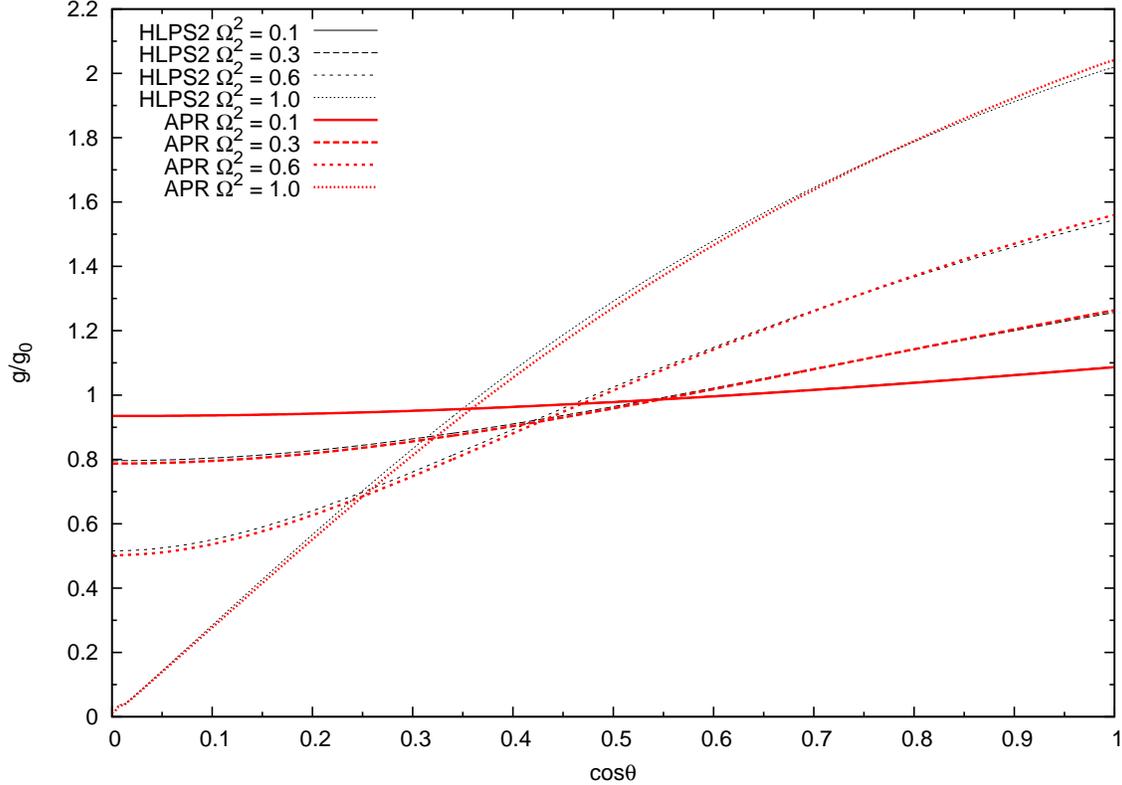}
\end{center}
\caption{Normalized effective acceleration due to gravity versus angular position on the star for a fixed compactness
ratio of $M/R = 0.195$. Plots for a few
different representative stellar models computed with EOS HLPS2 (black curves) and APR (red curves) are shown. 
Curves are labelled by 
the value of the square of dimensionless angular velocity. 
A non-rotating star would be 
represented by a horizontal line with $g/g_0 = 1.0$. All curves have the basic characteristic that the effective 
acceleration is smallest at the equator ($\cos\theta = 0$) and largest at the poles ($\cos\theta=1$).  
As the angular velocity increases, there is an increase in the deviation from unity.
Curves with the same value of $\bOmega^2$ have very similar shapes. 
See Table \ref{tab:HLPS2} for the values of the physical parameters for these models.
  }
\label{fig:gversusmu}
\end{figure}

\begin{figure}
\begin{center}
\plotone{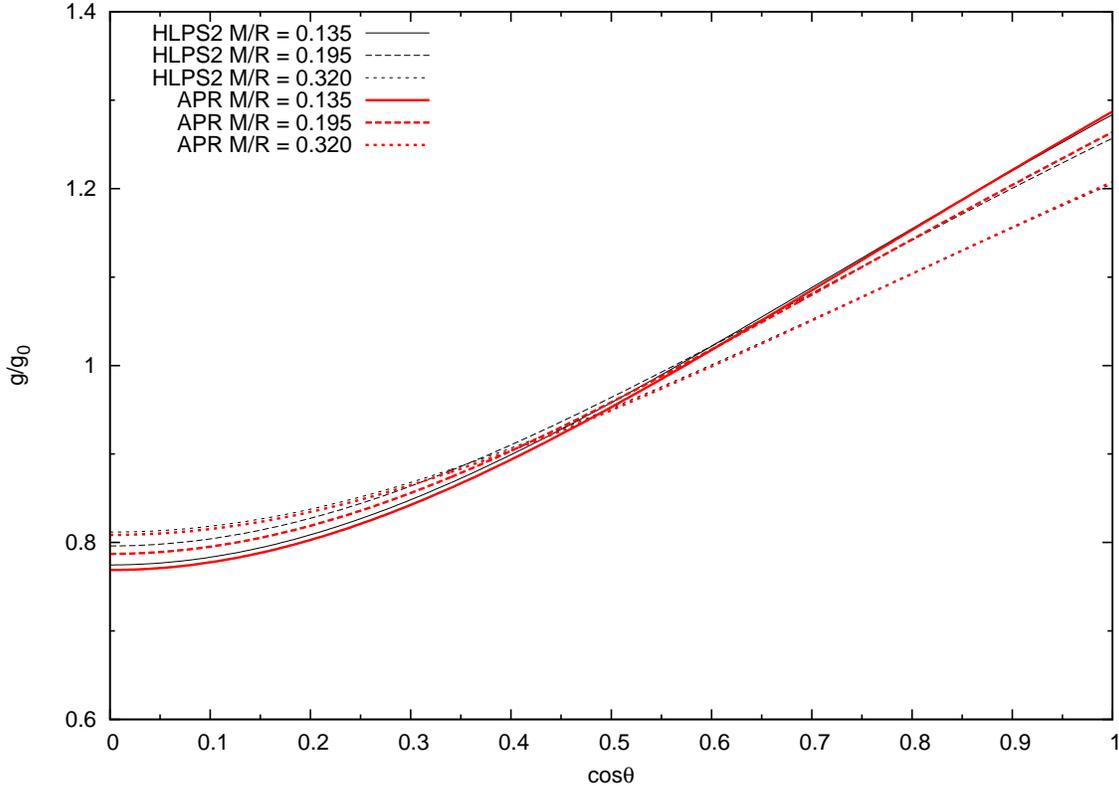}
\end{center}
\caption{Normalized effective acceleration due to gravity versus angular position on the star for a fixed 
angular velocity squared ratio $\bOmega^2=0.3$. Plots for a few
different representative stellar models computed with EOS HLPS2 (black curves) and APR (red curves) are shown. 
Curves are labelled by 
the value of the dimensionless compactness ratio $M/R$. 
As the compactness increases, the relative change in the acceleration from pole to equator decreases.
See Table \ref{tab:omega} for the values of the physical parameters for these models.
  }
\label{fig:gmoverr}
\end{figure}

The similarity between the effective gravity at other latitudes for models 
with differing EOS but the same values of $M/R_e$ and $\bOmega^2$
are shown in Figures \ref{fig:gversusmu} and \ref{fig:gmoverr}.
In Figure \ref{fig:gversusmu} we show plots of the dimensionless effective surface gravity
 versus $\cos(\theta)$ on the surface of the star for the models with a fixed value of $M/R = 0.195$ 
as listed in Table \ref{tab:HLPS2}. 
The equator corresponds to $\cos(\theta)=0$ while the spin axis corresponds to $\cos(\theta)=1$. 
Due to the choice of normalization, a non-rotating star's acceleration due to gravity will be
represented by a horizontal line at $g/g_0 = 1$. As the angular velocity increases, the curves 
deviate more from a horizontal line. In the extreme cases of the models that are just below the mass
shed limit, the effective acceleration is very close to zero at the equator as would be expected by
the definition of the mass-shed limit.

The universal behaviour of the effective acceleration can be seen in Figure \ref{fig:gversusmu}. Consider two curves with
the same line style. These two curves represent models computed with different EOS and different values of mass, radius
and spin frequency. However, the two curves have the same value of the dimensionless parameters $M/R_e$ 
and $\bOmega^2$. In the cases of the two lowest values of $\bOmega^2 = 0.1, 0.3$, the curves are almost 
indistinguishable. For the two higher spin models, the difference between the two curves can be seen by eye, but is 
still small. If models computed with other EOS but with the same values of the two dimensionless parameters were plotted, their 
effective acceleration curves would also be very similar to those chosen for this plot.

A similar comparison for stars with fixed $\bOmega^2$ is shown in Figure \ref{fig:gmoverr}. The physical
parameters for stellar models with $\bOmega^2=0.3$ and varying values of $M/R_e$ are shown in Table \ref{tab:omega}.
A range of compactness ratios allowing reasonably low-mass stars ($M/R=0.135$) to models close to the maximum
mass limit ($M/R=0.320$) are shown.  The dependence
of the effective acceleration on $M/R$ is not as strong as the dependence on $\bOmega^2$. In Figure \ref{fig:gmoverr},
two curves for stars with the same value of $M/R$ and $\bOmega^2$ but different EOS are shown with the same line style.
The difference between the two EOS curves can be seen by eye (due to the scale of the axes on this plot). The difference in curves due to changes in EOS (for the same $M/R$) is smaller than the difference in curves due to changing the value of $M/R$ (keeping EOS fixed), 
especially at high latitudes. Similar curves will result if different EOS are used.

\section{ Slow Rotation Approximation for the Effective Gravity}
\label{s:slow}

The slow rotation approximation for the effective gravity can be computed by 
substituting the slow rotation expansion for the metric, equations (\ref{eq:nu_0} - \ref{eq:zeta})
into the relativistic equations for the acceleration, (\ref{eq:vel} - \ref{eq:g_butter}) and 
keeping terms up to and including $\bigo(\bOmega^2)$. This procedure can be simplified by 
noting that the angular component of the acceleration vector, $a_\theta = \bigo(\bOmega^2)$,
while the radial component has both terms that are order unity and second order in $\bOmega$.
As a result, equation (\ref{eq:g_butter}) reduces to $g = e^{\nu-\zeta} a_{\rb} + \bigo(\bOmega^4)$.
In addition, the definition of the velocity in equation (\ref{eq:vel}) is first order in $\Omega$, so 
the first order approximation
\be
V_1 = r (1-\frac{2M}{r})^{-1/2} \Omega \sin\theta \left(1-\frac{\omega}{\Omega}\right),
\ee
where $V = V_1 + \bigo(\bOmega^3)$ can be used in equation (\ref{eq:a_butter}). It should
be remembered that $\omega/\Omega$ is of order unity. This allows the expansion of the
effective gravity to 2nd order in $\bOmega$,
\begin{eqnarray}
g &=& e^{\nu_0-\zeta_0} \left(1 + \bOmega^2(\nu_2 - \zeta_2) + \bigo(\bOmega^4)\right) \times 
\left( \frac{d \nu_0}{d \rb} + \bOmega^2 \frac{d \nu_2}{d \rb} - V_1 \frac{\partial V_1}{\partial \rb}  + \bigo(\bOmega^4) \right)\\
&=& e^{\nu_0-\zeta_0} \left(  
\frac{d \nu_0}{d \rb} - V_1 \frac{\partial V_1}{\partial \rb} 
+ \bOmega^2 \frac{d \nu_2}{d \rb} + \bOmega^2 \frac{d \nu_0}{d \rb} (\nu_2 - \zeta_2)  + \bigo(\bOmega^4)
\right).
\end{eqnarray}
This then
suggests the convenient, although not coordinate invariant, split of the effective gravity into 
\be
g =
 g_0 + a_c + a_q + a_o,
\ee
where $g_0$ is the acceleration due to gravity on the surface of a non-rotating star with the same mass and equatorial radius as the spinning star, $a_c$ is the centrifugal acceleration, $a_q$ is the correction due to mass-energy and pressure quadrupoles, and $a_o$ is the correction due to the oblate shape of the star. 

The non-rotating term is easily found to be
\be
g_0 = e^{\nu_0 - \zeta_0} \frac{d \nu_0}{d \rb} = \frac{M}{R_e^{2}}\frac{1}{(1-\frac{2M}{R_e})^{1/2}},
\ee
where equations (\ref{eq:nu_0} - \ref{eq:omega_0}) are used.

The effective centrifugal acceleration is defined by
\begin{eqnarray}
a_c &=& - e^{\nu_0 - \zeta_0} V_1 \frac{\partial V_1}{\partial \rb} \label{ac1}\\
&=& - g_0 \bOmega^2 \sin^2\theta \left(1-\frac{\omega}{\Omega}\right) 
	\left( \frac{(1-3M/R_e)}{(1-2M/R_e)} \left( 1 - \frac{\omega}{\Omega}\right) 
	- {\left(1-\frac{2M}{R_e}\right)^{-1/2}} \left(\frac{ \rb \partial \omega}{\Omega \partial \rb} \right) \right)\label{ac2},
\end{eqnarray}
where some algebra is needed to go from definition (\ref{ac1}) to (\ref{ac2}). The factor $(1-3M/R_e)$ has been derived and discussed in detail by \cite{Abram}, while the terms containing $\omega$ and $\partial_{\rb} \omega$ are
equivalent to the 2nd and 3rd terms appearing in the equation for the critical force calculated by \citet{LM95} (see their equation (36)). Equation (\ref{ac2}) is exact for all values of $R$, but it is possible to do a post-Newtonian expansion simplify it. We have found that it is necessary to keep terms of order $x^2 = (M/R_e)^2$ in order to capture the correct dependence of the centrifugal term for the most compact stars. To this order, the centrifugal term is
\be
a_c = - g_0 \bOmega^2 \sin^2\theta \left( 1 + x \left( -1 + 2 i \right)
	+ x^2 
\left(-2 + 4 i - 8 i^2\right)  + \bigo(x)^3 \right),
\ee
 where the dimensionless moment of inertia function $i(x,\bOmega)$ defined in equation (\ref{eq:inertia}) has been used. 

The quadrupole correction term is
\be
a_q = e^{\nu_0-\zeta_0} \left(  
 \bOmega^2 \frac{d \nu_2}{d \rb} + \bOmega^2 \frac{d \nu_0}{d \rb} (\nu_2 - \zeta_2) 
\right).
\ee
Substituting in the slow rotation expansions (\ref{eq:nu_0} - \ref{eq:zeta}) we find that the 
quadrupole correction term is
\be
a_q = g_0 x^2 \left( q P_2(\cos\theta) 
\left(3\sqrt{1-2x} -x\right)
 \left(\frac{R_e}{\bar{R}_e}\right)^3 - 2 \beta  \cos^2\theta \left(\frac{R_e}{\bar{R}_e}\right)^2\right),
\ee
where $\bar{R}_e$ is the equatorial radius of the star in isotropic coordinates. 
Since $q$ is negative, this leads to an increase in the acceleration on the equator and a decrease at the poles. 

The oblateness correction term corrects for the oblate shape of the star. In the slow rotation approximation, it is
sufficient to define this as the difference between the acceleration term for a non-rotating star evaluated at any angle $\theta$ from the pole and the equatorial acceleration,
\begin{eqnarray}
a_o &=& \frac{M}{R^2(\theta)} \frac{1}{\sqrt{1-2M/R(\theta)}} - \frac{M}{R_e^2} \frac{1}{\sqrt{1-2M/R_e}} \label{eq:goblate}\\
&=& \frac{M}{R_e^2} \frac{1}{\sqrt{1-2M/R_e}} \left( 
\left( \frac{R_e}{R(\theta)} \right)^2\left( 1 + \frac{2M(1/R(\theta) - 1/R_e)}{1-2M/R(\theta)}
\right)^{1/2} -1
 \right)\\
&=&  \frac{M}{R_e^2} \frac{1}{\sqrt{1-2M/R_e}} \frac{(R_e-R(\theta))}{R_e} 
\left( 2 + \frac{2M}{R_e} \times \frac{1}{1-2M/R_e} \right) + O(\bOmega^4).
\end{eqnarray}
Since the radial distance to the surface, $R(\theta)$ is always smaller than the equatorial
 radius, the oblateness correction increases the effective acceleration at latitudes above the
 equator, as would be expected.

This split of the acceleration into quadrupole and oblate terms is not coordinate invariant, since coordinate transformations that mix the radial and polar angular coordinate will change the size of the oblate corrections
and the quadrupole correction terms. However, the combination of terms is coordinate invariant.

Consider the relative magnitudes of the different contributions to the surface gravity. The leading order
correction due to the "centrifugal" force is $~ - g_0 \bOmega^2 \sin^2\theta$ while the leading order 
contribution from the oblate shape is $ ~ 1.6 g_0 \bOmega^2 \cos^2\theta$, so the oblate shape introduces an increase to the effective acceleration at the pole that is similar in magnitude to the decrease in the effective acceleration at the equator.  Since the leading order contribution to the quadrupole moment is $q \sim -  0.1 \bOmega^2 (R/M)^2$, the
acceleration correction due to the mass quadrupole is $ ~ - 0.3 g_0 \bOmega^2  P_2(\cos\theta)$, which is about an order of 
magnitude smaller than the oblateness and centrifugal corrections. The correction terms arising from the $\beta$ terms are
of the next post-Newtonian order and are negligible. 

In the slow rotation limit for the effective gravity, each term depends only $\cos^2\theta$ or $\sin^2\theta$. This allows the simple 
approximation
\be
\frac{g(\theta)}{g_0}  = \left( 1 + c_e \bOmega^2 \sin^2\theta + c_p \bOmega^2 \cos^2\theta\right),
\label{eq:slowaccel}
\ee
where the expansion coefficients for the equatorial and polar accelerations are
\be
\frac{g_{(e,p)}}{g_0} = 1 + c_{(e,p)} \bOmega^2 = 1 +  (c_{0(e,p)} + c_{1(e,p)} x)\bOmega^2.
\label{eq:gep}
\ee
The shapes of the $g/g_0$ versus $\cos\theta$ curves for slow rotation ($\bOmega^2=0.1$) shown in Figure \ref{fig:gversusmu} are very well approximated by 
the functional form given by Equation (\ref{eq:slowaccel}). Since there are only two independent functions of $\cos\theta$ in 
Equation (\ref{eq:slowaccel}) the values at all latitudes are given once the values at the pole and equator have been determined.
We computed a wide range of stellar models with $\bOmega^2\le 0.1$, as described in Section \ref{s:para}. For each model
the effective acceleration due to gravity at the equator and at the pole has been computed. An empirical least-squares fit for the dependence of
the effective accelerations on the dimensionless parameters yields the best-fit functional form shown in Equation (\ref{eq:gep}). The
 best-fit values for the expansion parameters are shown in Table \ref{tab:gcoeff}.

\begin{deluxetable}{llrrr}
\tablecaption{Expansion Coefficients for the Effective Gravity for Slow Rotation. See Equation (\ref{eq:gep}) for definitions.}
\tablehead{ \colhead{Quantity} & \colhead{Equation}& 
 \colhead{$c_{0(e,p)}$} & \colhead{$c_{1(e,p)}$} & \colhead{$\sigma$}   }
\startdata
$c_e \bOmega^2$&  $ (c_{0(e)} +c_{1(e)} x)\bOmega^2 $  &   $-0.791 \pm 0.2\%$ &   $0.776 \pm 0.7\%$  & 0.001 \\
$c_p \bOmega^2$& $  (c_{0(p)} +c_{1(p)} x)\bOmega^2 $ &    $ 1.138 \pm 0.1\%$ & $-1.431  \pm 0.3\%$ & 0.001 \\
\enddata
\label{tab:gcoeff}
\end{deluxetable}

\section{Rapid Rotation}
\label{s:rapid}

For rapidly rotating stars we expect a few changes from the simple formula (\ref{eq:slowaccel}). At higher rotation rates ($\bOmega^2\ge 0.1$), the star's surface will be more deformed, and higher order multipole moments in the star's gravitational field will become important. As seen in 
Figures \ref{fig:gversusmu} and \ref{fig:gmoverr}, the dependence on angle is somewhat more complicated than suggested by
the slow rotation formula (\ref{eq:slowaccel}) and the deviation from a constant value of acceleration over the surface of the
star is more pronounced.  We have found a better
fit for the effective acceleration at larger angular velocities can be found by adding a term linear in $\cos\theta$ to the slow rotation fit. Since
there are three independent angular functions ($\cos\theta$, $\cos^2\theta$, and $\sin^2\theta$), empirical fits can
be found by computing the acceleration at three points on the surface. We have chosen to do this calculation at the equator, the pole and at an angle of 60 degrees from the pole. This leads to the following empirical formula for the effective 
acceleration due to gravity on the surface of a rapidly rotating neutron star,
\be
\frac{g(\theta)}{g_0}  =  1 + (c_e \bOmega^2 + d_e \bOmega^4 + f_{e} \bOmega^6) \sin^2\theta + 
(c_p \bOmega^2 + d_p \bOmega^4  + f_p \bOmega^6 - d_{60}\bOmega^4) \cos^2\theta +  d_{60}\bOmega^4 \cos\theta.
\label{eq:rapidg}
\ee

The values of the coefficients in Equation (\ref{eq:rapidg}) shown in Table (\ref{tab:gcoeff-rapid})
were found by computing the values of the effective acceleration at three latitudes 
($\cos\theta$ = 0, 0.5, 1) for a large range of stellar models with $\bOmega^2 \le 1.0$. The least-square fits for the coefficients
were performed while imposing the constraint that the coefficients $c_{e,p}$ have the same values given by the slow-rotation
fits in Equation (\ref{eq:gep}). Allowing the values of $c_{e,p}$ to vary in the fits did not significantly improve the quality of the fits.

\begin{deluxetable}{llrrrl}
\tablecaption{Expansion Coefficients for the Effective Gravity for Rapid Rotation. See Equation (\ref{eq:rapidg}) for definitions.}
\tablehead{ \colhead{Quantity} & \colhead{Equation}& 
 \colhead{$d_{0(e,p,60)}$} & \colhead{$d_{1(e,p,60)}$} & \colhead{$d_{2(e,p,60)}$}   & \colhead{$\sigma$}}
\startdata
$d_e\bOmega^4$&  $ (d_{1(e)}x +d_{2(e)} x^2)\bOmega^4 $ & &  $-1.315 \pm 0.3\%$  &  $2.431 \pm 0.5\%$ &  0.009\\
$f_e \bOmega^6$&  $ f_{1(e)}x\bOmega^6 $ & & $-1.172 \pm 0.3\%$   && 0.009\\
$d_p\bOmega^4$& $  (d_{1(p)}x +d_{2(p)} x^2)\bOmega^4 $ & &   $0.653\pm 1.2\%$  & $-2.864 \pm 0.7\%$ & 0.016 \\
$f_p \bOmega^6$& $ f_{1(p)}x \bOmega^6$ &  & $0.975 \pm 0.7 \%$ && 0.016 \\
$d_{60}\bOmega^4$&$(d_{1(60)} x + d_{2(60)} x^2)\bOmega^4 $ & & $13.47\pm 0.2\%$  & $-27.13 \pm 0.2\%$ & 0.007   \\
$f_{60} \bOmega^6$&$f_{0(60)}\bOmega^6$ & $1.69 \pm 0.2\%$ &&&0.007\\
\enddata
\label{tab:gcoeff-rapid}
\end{deluxetable}

\begin{figure}
\begin{center}
\plotone{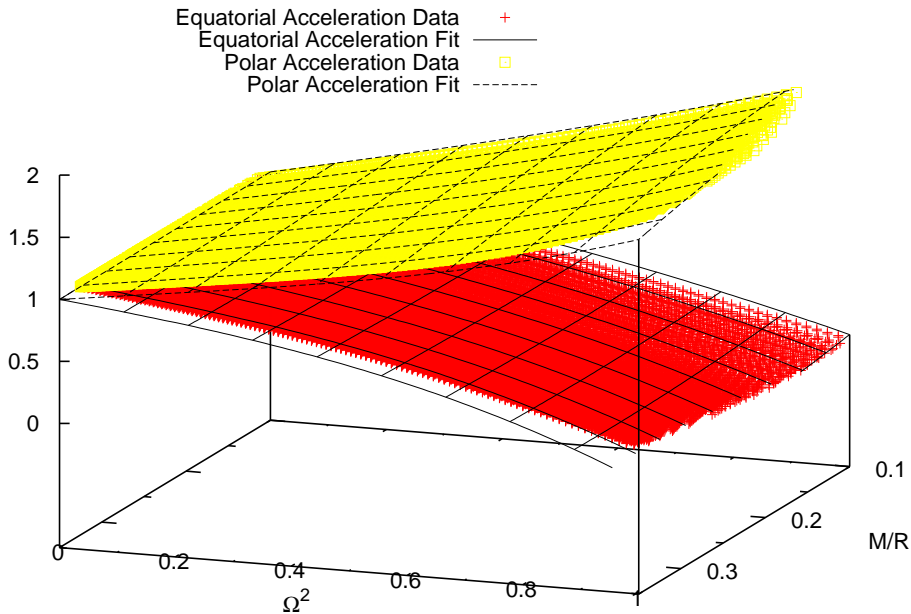}
\end{center}
\caption{Normalized effective acceleration due to gravity as a function of $M/R$ and $\bOmega^2$.
Each point corresponds to a neutron star model computed with one of the EOS listed in Section 3.
The lower surface defined by red crosses are values of acceleration at the equator,
while the upper surface defined by yellow squares denote values of the acceleration at the
pole.  
The solid and dashed black surfaces corresponds to the best fit function given in Equation (\ref{eq:rapidg}), evaluated at
$\cos\theta=0$ and $\cos\theta=1$ respectively. 
  }
\label{fig:equat}
\end{figure}

In Figure \ref{fig:equat}, the dimensionless effective gravity for approximately 39,000 neutron star 
models constructed with the EOS listed in Section 3 are shown with points that form two surfaces. 
The  red crosses 
denote values of acceleration at the equator while the yellow squares denote values of acceleration at the poles. 
Each set of points forms a surface (upper surface demarked by squares, lower surface demarked by crosses) that is independent of the EOS.
 The best fit to these surfaces, Equation (\ref{eq:rapidg})
 is plotted as a solid (for equatorial acceleration) or dashed (for polar acceleration) gridded surface.

\section{Discussion and Conclusions}
\label{s:final}

In this paper we computed the effective acceleration due to gravity as measured in the rotating frame on the surface of a 
rapidly rotating relativistic neutron star. We find that the effective gravity can be written as the simple function
$g(\theta) = c(M/R_e, \Omega^2 R_e^3/(GM),\theta) g_0$, where $g_0 = GM/(R_e^2\sqrt{1-2GM/R_ec^2})$ is the
acceleration due to gravity on the surface of a non-rotating relativistic star. The dimensionless function $c$ has a 
universal form in that it is almost independent of the neutron star's equation of state, and only depends
on the dimensionless parameters $x=M/R_e$ and $\bOmega^2 = \Omega^2 R_e^3/(GM)$ as well as latitude on the
star's surface. The dimensionless function $c$ is expanded in the slow rotation limit to bring about an intuitive understanding
of the contributions to the effective gravity arising from the centrifugal force, the oblate shape and the quadrupole
moment of the gravitational field. In addition, we provide an empirical fit to the function $c$ for rapid rotation. As
expected, the effective centrifugal force decreases the effective gravity, while the centrifugal flattening of the star
into an oblate shape increases the effective gravity at the poles. In addition, the quadrupole moment of the 
gravitational field increases the effective gravity at the equator and decreases it at the poles, however the 
quadrupole correction is a small correction to the main effects due to the effective centrifugal force.
For all rotation rates and compactness ratios, the
increase in the effective gravity at the poles is of the same order of magnitude as the decrease in the effective 
gravity at the equator.

The slow rotation limit ($\bOmega^2 \le 0.1$) is a good approximation for neutron stars and pulsars with spin frequencies up to 
about $600$ Hz. This statement is EOS dependent, but it would take a very low mass and large radius (say $1.0 M_\odot$ 
and 15 km) to bring $\bOmega^2$ up to 0.3 which is the maximum value of the spin parameter that the slow rotation limit 
can be applied to. For the observed ms-period pulsars and accreting neutron stars, the slow rotation approximation for the
effective acceleration, Equation (\ref{eq:slowaccel}), is valid. Using this approximation, the difference between the 
effective acceleration at the pole and equator is
\be
{g_p - g_e} = g_0 \bOmega^2 (|c_p| + |c_e|) = g_0 \bOmega^2 ( 1.93 - 2.2 M/R_e ).
\ee 
Similarly, the fractional change (compared to the equator) is
\be
\frac{g_p - g_e}{g_e} = \bOmega^2( |c_p| + 2 |c_e|) = \bOmega^2 (2.72 - 3.0 M/R_e),
\ee
which is close to 3 times larger than the estimate $(g_p-g_e)/g_e \sim \bOmega^2$
used by many authors (for example, see \citet{Watts}).
For one of the representative stellar models with $\bOmega^2=0.1$ and $M/R_e = 0.195$ given in Table \ref{tab:HLPS2},
the fractional change in the effective acceleration with respect to the equator is 0.2, or a 20\% change.

Most atmosphere models include a dependence on the effective acceleration due to gravity as one of the parameters.Typically,
the dependence is on the logarithm of the surface gravity. It is useful then to compute the difference in the logarithm of
the surface gravity at the pole and at the equator,
\be
\log(g_p) - \log(g_e) = \bOmega^2 (|c_p| + |c_e|) .
\ee
For the same representative models, this is a difference of 0.15 in dex.  Although this seems like a small change, 
this is the same change in acceleration one would find by changing the mass of the star by 15\%. Changes of 
this order of magnitude make small but visible changes to the flux predicted by Hydrogen atmosphere models
\citep{Heinke06}, which are used to model  X-ray transients and X-ray ms-pulsars.
 The X-ray pulsars studied by \citet{Bog} are rotating slowly enough ($\nu_\star\sim 200$ Hz) that the latitudinal dependence of the
effective acceleration is not important. However, the X-ray transients (such as EXO 0748-676 with a spin frequency of 552 Hz) 
are spinning rapidly enough that the change in gravity over latitude may affect the interpretation of observations such
as those by \citet{Degen}. Similarly, atmosphere models \citep{Sul} for neutron stars undergoing Type I X-ray bursts 
also depend the effective gravity.

The Type I X-ray bursts observed in some X-ray binaries are caused by unstable nuclear burning on the 
surfaces of neutron stars (see \citet{Galloway} for a review). There are at least 10 bursting neutron stars
with spin frequencies of 500 Hz or more \citep{Watts}, which is fast enough that the change in 
gravitational acceleration over the surface is a significant factor. In early studies of Type I X-ray bursts
\citep{Spitk}, it was argued that the ignition would take place at the equator since that is where the 
effective acceleration due to gravity is smallest. However, \citet{Cooper} showed that the critical 
mass-accretion rate, which is the largest rate that allows unstable nuclear burning, varies with
the effective gravity as $\dot{M} \sim g(\theta)^{3/2}$. This leads to range of mass accretion
rates that are too high to allow ignition at the equator, but allow ignition at higher latitudes due
to the higher surface gravity. Since the fractional change in the effective acceleration is 
about 2.5 times larger than that estimated by \citet{Cooper}, the range of mass accretion
rates that could lead to high latitude ignition is larger, roughly 20-30\% of the 
critical mass accretion rate at the equator. \citet{Maurer} have shown that there is
good evidence that off-equatorial
burst ignition is taking place in the burster 4U 1636-536 and may be relevant in
other Type I X-ray bursters.

The Eddington luminosity at the surface of a neutron star 
can be written 
\be
L_{Edd} = L_0 \frac{ R^2(\theta) (1-2M/R(\theta)) g(\theta)}{M},
\label{eq:edd}
\ee
where $L_0$ is the regular expression for the Eddington luminosity (see \citet{Ozel2013} for a review) in flat space
\be
L_0 = \frac{ 8\pi G M m_p c}{(1+X)\sigma_T},
\ee
which is independent of the star's radius. 
In Equation (\ref{eq:edd}), the factor $(1-2M/R)$ corresponds to the gravitational redshift of the 
light, and we are not considering (at this time) the extra Doppler shift due to rotation.
The dependence of the Eddington luminosity on the effective gravity
leads directly to a reduction of the Eddington limit at the star's equator and an increase at the poles. 
The Eddington limit at the equator is
\be
L_{Edd,e} = L_0 ( 1 - 2M/R_e )^{1/2} (1 - |c_e| \bOmega^2),
\ee
while at the pole it is
\be 
L_{Edd,p} = L_0 ( 1 - 2M/R_p )^{1/2},
\ee
where we are approximating the effective gravity at the pole by just the oblate term given by Equation (\ref{eq:goblate}). The 
ratio of the gravitational redshift terms at the pole and equator is of order $\bigo(\bOmega^2 \times M/R)$ 
so the term proportional to $c_e$ is the most important correction due to rotation.
For stars with a rotation parameter near $\bOmega^2=0.1$, we then expect 
a difference in the Eddington limit of up to 10\% between the poles and equator.
It should be noted that \citet{LM95} carefully considered the effect of rotation
on the Eddington limit and found no changes. However, their analysis was for
stars rotating more slowly than the ones that we consider, so there is no 
conflict in our results.

During a photospheric radius expansion (PRE) burst, the flux generated by the Type I 
X-ray burst is large enough to exceed the Eddington limit, causing the photosphere to 
expand to many times the radius of the star \citep{Lewin}.  The change in the Eddington 
limit given above is only valid when the atmosphere is in hydrostatic equilibrium and
co-rotating with the rest of the star. As the photosphere expands, if the material 
continues to co-rotate, it will quickly reach the mass shed limit which will have the 
effect of taking the material in the photosphere out of co-rotation. At the end of
the burst, the flux is reduced to below the Eddington limit and the photosphere
"touches down" onto the surface of the star, regaining hydrostatic balance
and co-rotation. Since the Eddington limit is largest at the poles and smallest
at the equator, we expect that touch-down first occurs at the poles and then
quickly moves down to the equator as the atmosphere cools. The relevant
Eddington flux that should be equated with the asymptotic "touch-down"
flux is the Eddington flux at the equator.  This will introduce
correction terms for  the masses 
and radii derived (for example see \citet{Ozel12})
from measurements of PRE-bursts for the most rapidly
rotating stars, such as KS 1731-260 which has a spin of 524 Hz. 
A full treatment of the corrections arising from rapid rotation on
the masses and radii derived from PRE burst is beyond the scope
of this paper and requires an examination of the apparent area
of the bursting star \citep{Baubock2012} as well as raytracing to 
find the mapping between the locally produced flux and the flux 
measured at infinity.

\acknowledgments
This research was supported by a grant from NSERC. 
SMM thanks Michi Baubock, Feryal Ozel and Dimitrios Psaltis
for helpful discussions.
SMM also thanks the Steward Observatory for hosting her 
stay at the University of Arizona during her sabbatical visit.

\end{document}